\documentclass[twocolumn,showpacs,preprintnumbers,amsmath,amssymb]{revtex4}

\usepackage{graphics}
\usepackage{dcolumn}
\usepackage{bm}

\begin{document}

\title{Step-like magnetization in a spin-chain system: Ca$_3$Co$_2$O$_6$}

\author{Yuri B. Kudasov}
\email{kudasov@ntc.vniief.ru}
\affiliation{Institut de Physique et Chimie des Mat\'eriaux de Strasbourg, 67034, 
Strasbourg, France\\
Russian Federal Nuclear Center - VNIIEF, Sarov, 607188, Russia }

\date{\today}
\begin{abstract}
Due to a ferromagnetic in-chain coupling between Co$^{3+}$ ions at trigonal sites, chains 
Co$_2$O$_6$ are considered as large rigid spin moments. The antiferromagnetic Ising model on 
the triangular lattice is applied to describe an interchain ordering. An evolution of 
metastable states in a sweeping magnetic field is investigated by the single-flip technique. 
At the first approximation two steps in the magnetization curve and a plateau at $1/3$ of 
the saturation magnetization are found. Four steps in magnetization are determined in 
high-order approximations in agreement with experimental results.
\end{abstract}

\pacs{75.25.+z, 75.30.Kz, 75.50.Ee}

\maketitle

Among other spin-chain compounds, Ca$_3$Co$_2$O$_6$ has drawn recently considerable 
attention to their complex magnetic behavior \cite{hardy1, drillon, hardy2, hardy3, 
maignan}. The most intriguing feature observed in Ca$_3$Co$_2$O$_6$ is a step-like shape of 
the magnetization curve \cite{drillon, hardy2, maignan}. The number of the steps in the 
curve depends strongly on a sweep rate of the external magnetic field and temperature 
\cite{drillon, hardy2, maignan}. Two steps become apparent in the temperature range from 12 
K to 24 K \cite{maignan}. The first step takes place at the zero magnetic field. Then the 
magnetic moment remains constant at about $1/3$ of the full magnetization up to the magnetic 
field of 3.6 T where the second step occurs to the fully magnetized FM state.  At least four 
equidistant steps are clearly visible below 10 K at a very low sweep rate. They are 
accompanied by a sizeable hysteresis. Similar phenomena were observed in other spin-chain 
compounds, e.g. Ca$_3$CoRhO$_6$ \cite{nijtaka1, nijtaka2}.

The structure of Ca$_3$Co$_2$O$_6$ consists of Co$_2$O$_6$ chains running along the $c$ 
axis. The Ca ions are situated between them. The chains are made up of alternating, 
face-sharing CoO$_6$ trigonal prisms and CoO$_6$ octahedra. The crystalline electric field 
splits the energy level of Co$^{3+}$ ions into the high-spin ($S=2$) and low-spin ($S=0$) 
states.
The Co$^{3+}$ ions situated in the trigonal environment (CoI) are in the high-spin state and 
the octahedral Co sites (CoII) occurs in the low-spin state. In the last case the energy 
difference between the low-spin and high-spin states is very small and a tiny fraction of 
CoII sites is reported to be in the high-spin state. The crystalline electric field leads 
also to a very strong Ising-like anisotropy at the CoI sites. The chains form triangular 
lattice in the $ab$ plane that is perpendicular to the chains. An in-chain exchange 
interaction between magnetic CoI ions through the octahedra with non-magnetic CoII ions is 
ferromagnetic (FM).  It causes the in-chain FM ordering of CoI ions at about 40 K. The 
interchain interaction is antiferromagnetic (AFM) and much weaker than the in-chain one. A 
partial AFM order of chains appears at 24 K. A weak feature concerned most probably with a 
transition in a new interchain order was also observed at around 12 K. This scenario of 
magnetic interactions in Ca$_3$Co$_2$O$_6$ is consistent with results of x-ray photoemission 
spectroscopy \cite{x-ray}, neutron scattering \cite{neutron}, magnetization  and specific 
heat measurements \cite{hardy1, drillon, hardy2, hardy3, maignan,martinez}, nuclear magnetic 
resonance \cite{nmr}, and theoretical calculations of indirect interactions between CoI 
sites \cite{fresard}.

The model presented in Ref. \cite{drillon} deals with an in-chain structure of 
Ca$_3$Co$_2$O$_6$ and magnetization dynamics explained in terms of the quantum tunnelling. 
In the this Letter, we develop a new model for a description of the step-like magnetization 
in Ca$_3$Co$_2$O$_6$ at low temperatures, shifting the stress on the interchain magnetic 
order. The strong FM in-chain coupling makes it possible to consider a Ca$_2$O$_6$ chain as 
a large rigid spin formed by CoI ions. There are only two projections of the chain spin onto 
the $c$ axis due to the strong Ising-like anisotropy. Including the AFM coupling between the 
nearest-neighbor chain spins we arrive to the Ising Hamiltonian on the triangular lattice
\begin{equation}
H=J\sum_{<ij>}{\sigma_i^z \sigma_j^z}-B\sum_{i}{\sigma_i^z}
\label{Ising}
\end{equation}
where $\sigma_i^z=\pm{1}$ is the $c$-axis projection of the $i$-th chain spin, $J>0$ is the 
parameter of the AFM interchain coupling, $B$ is the magnetic field, $<ij>$ denotes the 
summation over all the nearest-neighbor pairs on the triangular lattice.

The strong dependence of the magnetization curve shape on the magnetic field sweep rate and 
temperature shows that the state of the system of the chain spins in the magnetic field is 
far from equilibrium. At the low sweep rate the system is rather in a metastable state than 
in the ground state. It is convenient to formulate necessary conditions of the metastability 
of the system in the following form  
\begin{equation}
\sigma_i^z h_i\leq 0
\label{meta}
\end{equation}
where 
\begin{equation*}
h_i = J\sum_{j(i)}{\sigma_j^z}-B
\end{equation*}
is the effective field for the $i$-th chain, $j(i)$ denotes summation over the 
nearest-neighbors of the $i$-th chain. We have used the unstrict inequality in 
Eq.(\ref{meta}) keeping in mind a strong degeneracy of partially ordered AFM arrangements on 
the triangular lattice. 
A transition from one metastable state to another occurs through exited states. We obtain 
$\sigma_i^z h_i > 0$ at least for one chain in a exited state by definition. Let $\Delta E_i 
= 2 \sigma_i^z h_i> 0$ be the excitation energy per CoI site of the $i$-th chain. It follows 
from this that the probability of an exited chain at low temperature $T$ is extremely small 
$\propto exp(- N \Delta E_i / T)$ since the number of CoI sites in the chain ($N$) is 
considered to be large. That is why, we should investigate an evolution of metastable states 
in the slowly sweeping external magnetic field assuming $T=0$ quench.

We perform the investigation of the evolution of the system using the single-flip technique 
that was applied earlier to nonequilibrium dynamics of the AFM Ising model on the the 
triangular lattice \cite{kim}.  In the terms of the effective field, the spin-flip 
probability $A$ is taken in the following form
\begin{eqnarray}
A=\left \lbrace \begin{array}{cc} 
0 & \text{ if }\sigma_i^z h_i < 0, \\ 
1 & \text{ if }\sigma_i^z h_i \geq 0. 
\end{array}\right.
\end{eqnarray}

In contrast to Ref.\cite{kim} where the spin-flip technique was applied to numerical Monte 
Carlo simulations, we investigate the evolution analytically. If a state under consideration 
is degenerate and different sequences of spin flips lead to different final states, one 
should take into account each possible sequence with equal probabilities. This assumption 
can be proven rigorously by Bogolubov's quasiaverage technique. That is a small auxiliary 
random field should be added to the Hamiltonian (\ref{Ising}) in order to lift the 
degeneracy. After that we take an average over a manifold of the auxiliary fields restoring 
equivalence of different chains and, then, let the amplitude of the auxiliary fields goes to 
zero. This approach is equivalent to the Monte Carlo technique in the limit of the large 
number of samples.

We take the ground state of the triangular lattice as an initial state of the chain lattice 
at $B=0$. The ground state of triangular lattice at $B=0$ is strongly degenerate and it is 
impossible to represent it in an explicit form \cite{wannier}. On the other hand, we can 
produce a set of approximations for the ground state. Since the entropy density for the 
ground state was calculated exactly by Wannier \cite{wannier} as 
\begin{eqnarray}
S = \frac{2}{\pi} \int^{\frac{\pi}{3}}_{0}{\ln (2 \cos \omega)} d\omega \approx 0.3383,
\end{eqnarray}
we are able to compare the entropy density of the approximated state to the exact value in 
order to control the precision of our approximation. Previously, Maignan \emph{et al}. 
\cite{maignan} performed a qualitative analysis of the magnetization curve starting with an 
initial state that consisted of alternating rows of 
spin-up and spin-down chains. The energy of such an arrangement equals the ground state 
energy but its statistical weight goes to zero in the limit of the infinite lattice. 
Therefore, arrangements of this type should be discarded \cite{wannier}.

\begin{figure}
\includegraphics{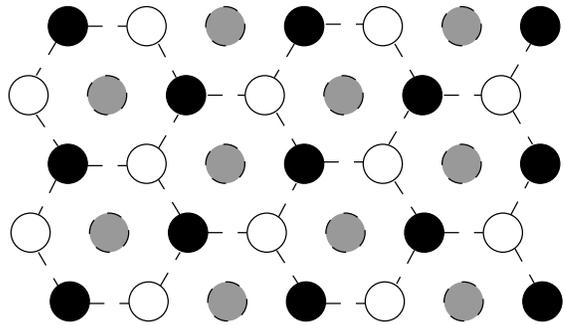}
\caption{\label{f1} The honeycomb magnetic structure. The black and white circles are 
spin-up and spin-down states, correspondingly, the gray circles can be either spin-up or 
spin-down states.}
\end{figure}

\begin{figure}
\includegraphics{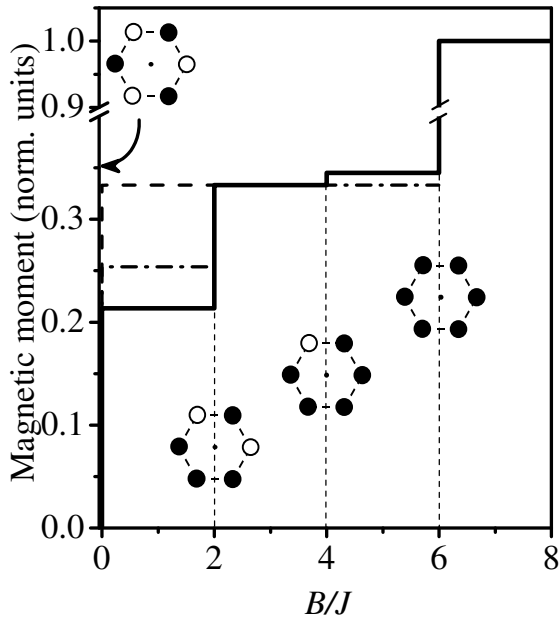}
\caption{\label{f4} The magnetic moment as a function of the dimensionless magnetic field 
$B/J$. The dash, dash-dot, and solid lines are the results of the first, second, and fourth 
approximations, correspondingly. The four nearest-neighbor configurations producing the 
critical spin-flip fields are shown. Note the break in the vertical axis.}
\end{figure}

The first approximation to the ground state of the AFM Ising model on the triangular lattice 
is the honeycomb structure shown in Fig.\ref{f1}. Two thirds of the total number of the 
chain spins are ordered in the AFM honeycomb structure whereas other chain spins placed in 
the centers of hexagons have arbitrary projections of chain spins onto $c$ axis. The entropy 
density of the first approximation equals $S_1=(1/3)ln 2\approx 0.231$. An arbirary small 
external magnetic field lifts degeneracy of chain spins in the centers of hexagons orienting 
them along the magnetic field ($\sigma_i^z=1$). We consider that the spin-up chains (the 
black circles) are directed parallel to the magnetic field. The grey circles become black in 
Fig.\ref{f1}. This causes a step at $B=0$ with the height of $1/3$ where the full 
magnetization is taken as unity. Then, one third of the spin chains remain spin-down or 
antiparallel to the field. Since they are surrounded by 6 spin-up chains this configuration 
is stable up to the critical magnetic field $B_C=6J$ where a transition to the 
fully-polarized FM state takes place. The magnetization curve for the first approximation is 
shown in Fig.\ref{f4}. It should be noticed that this curve is in an excellent agreement 
with the experimental data at the intermediate temperatures \cite{hardy1, drillon, hardy2, 
maignan}.

\begin{figure}
\includegraphics{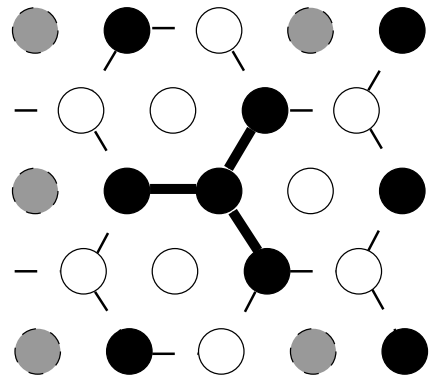}
\caption{\label{f2} An arrangement containing the tripod configuration. The tripod is shown 
in the solid line. The three white circles close to the center of the tripod have 4 spin-up 
(the black circles) and 2 spin-down (the white circles) nearest-neighbors.}
\end{figure}

We can improve the approximation used above including tripod configurations \cite{wannier}. 
Since the chains placed in the centers of hexagons have an arbitrary projections of the spin 
onto $c$ axis, it occurs that three of them neighboring with the same chain of the honeycomb 
sublattice are in the same state. Then the configuration shown in Fig.\ref{f2} can appear. 
The chain spin in the center of the tripod belongs to the honeycomb sublattice but, 
nevertheless, it can have an arbitrary $c$ projection of the chain spin. These states 
increase the entropy density up to $S=(5/12)ln2\approx 0.289$ \cite{wannier} and involve 
various types of configurations: isolated tripods and connected tripods. It is convenient to 
consider them
separately. After straightforward calculations we obtain the probability of an isolated 
tripod
\begin{equation}
P_2=(1/12)(1-1/2^3-1/2^4-1/2^5)^3 \approx 0.0397.
\end{equation}
We include this type of configurations in the second approximation for the initial state. 
The entropy density of the second approximation is $S_2 \approx 0.259$. 
Applying the external magnetic field we again obtain a step at the zero magnetic field but 
the chain spins in the centers of hexagons sharing joint corners with the tripod center 
remain in the spin-down state because they are surrounded by 4 spin-up and 2 spin-down 
nearest-neighbors. The height of the step ($\Delta M(B/J)$) can be expressed through 
probability of the isolated-tripod configuration as $\Delta M(0)=(1/3)-2P_2$. The spin 
chains in the centers of the tripod hexagons flip at the new critical magnetic field 
$B_C=2J$. While the magnetic fields increases further, the curve occurs at the $1/3$-plateau 
and coincides with the curve obtained in the first approximation (see Fig. \ref{f4}).

The third approximation is to include configurations with isolated pairs of connected 
tripods. These configurations change the heights of the steps but give no new features in 
the magnetization curve. The probability of the pair of isolated tripod is $P_3 \approx 
0.010$. The entropy density increases up to $S_3 \approx 0.273$. 

\begin{figure}
\includegraphics{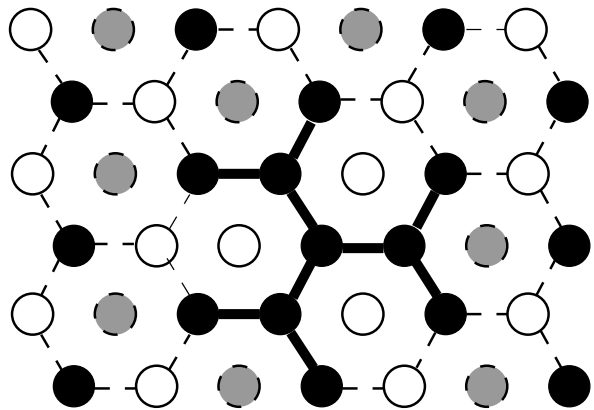}
\caption{\label{f3} Three tripods connected in the star arrangement. Three white circles 
close to the center of the star are surrounded by 5 spin-up and one spin-down 
nearest-neighbors.}
\end{figure}

An isolated configuration of three tripods connected as a star is taken into account in the 
framework of the fourth approximation (see Fig.\ref{f3}). The probability of this 
configuration is $P_4 \approx 0.0035$ and the entropy density in the fourth approximation is 
$S_4 \approx 0.280$. The key feature of this configuration is that there are chain spins 
that are surrounded by 5 spin-up and one spin-down nearest-neighbors. These chain spins 
remain stable up to the new critical magnetic field $B_C=4J$. To calculate the step of in 
the magnetization curve we have to calculate a number of various sequences of spin flips. 
The final value of the step is $\Delta M(4) \approx 0.008$ that is significantly smaller 
then $\Delta M(2) \approx 0.12$.

There is a variety of  more complex configurations then the tripod. However, it should be 
mentioned that further approximations should change the heights of the steps but they can 
not cause new features in the magnetization curve. As it follows from Eq.(\ref{meta}) there 
exist only four critical magnetic fields related to the four configurations of the 
nearest-neighbors shown in Fig.\ref{f4}.

The magnetization curve obtained in the fourth approximation reproduces the key features of 
the experimental data at the very low sweep rate, namely, the four equidistant steps in the 
magnetization curve. In contrast with our results, the third step in the experimental curve 
is much larger than the second one. This quantitative discrepancy can be eliminated in 
higher order approximations that can be investigated both analytically, by the technique 
used in the present Letter, or numerically, applying Monte Carlo simulation \cite{kim}. 
Approximations of higher orders are of importance for quantitative calculations of the 
magnetization curve because the calculation convergence for the magnetic moment is slower 
than that for the configuration probability. 

Few questions on the magnetization of Ca$_3$Co$_2$O$_6$ are still unclear. There were 
observed weak smeared features in the experimental curve at high magnetic fields above the 
last step. They can stem from the small fraction of CoII sites that are the high-spin state, 
because
they increase the chain spin and cause higher critical fields. The sizeable hysteresis also 
draws attention in the experimental curve. It depends drastically on the temperature and the 
magnetic field sweep rate. It should be mentioned that while the magnetic field sweeps down 
and crosses the highest critical field $B_C=6J$ the chain spins flip down at random and the 
system occurs in a new state that is different from that at the sweeping-up process. 

In conclusion, we have developed a new model for the step-like magnetization of 
Ca$_3$Co$_2$O$_6$ spin-chain compound. It can be applied also to other spin-chain compounds 
with the triangular lattice of chains, e.g. Ca$_3$CoRhO$_6$. Due to the in-chain FM coupling 
between Co$^{3+}$ ions at trigonal sites, Co$_2$O$_6$ chains are considered at low 
temperatures as a large rigid spins with the strong Ising-like anisotropy. The AFM Ising 
model on the triangular lattice is applied to the system of rigid FM-ordered chains. The 
crucial point of the model is the supposition that the system is out of equilibrium, because 
the dependence of the magnetic moment on the magnetic field in the ground state of the AFM 
Ising model on the triangular lattice is smooth with the exception of the step at the zero 
magnetic field \cite{mattis}. For the honeycomb AFM structure two steps were found in the 
theoretical magnetization curve in excellent agreement with experimental data at the 
intermediate temperatures. At higher approximations four equidistant steps were determined 
in accordance with experimental curves at the low temperatures and very low magnetic field 
sweep rate. The results obtained in the present Letter and the model of Ref.\cite{drillon} 
can be regarded as two limiting cases. The first deals with the nonequilibrium interchain 
ordering assuming the chain spins to be rigid. An in-chain fragmentation and the quantum 
tunnelling of the magnetic moment of the fragments are investigated in the second case 
totally neglecting the interchain ordering. 

I am grateful to C. Demangeat and M. Drillon for fruitful discussions and hospitality during 
my stay at Institut de Physique et Chimie des Mat\'eriaux de Strasbourg en Unit\'e mixte de 
recherche.

This study was partially supported by the INTAS grant (03-51-4778) "Hierarchy of scales in 
magnetic nanostructures".

\end{document}